\documentclass[submission,copyright,creativecommons]{eptcs}
\providecommand{\event}{TERMGRAPH} %
\usepackage[utf8]{inputenc}
\usepackage{tikz}
\usetikzlibrary{arrows}
\usetikzlibrary{positioning}

\title{Efficient Completion of Weighted Automata}
\author{Johannes Waldmann
  \institute{Hochschule für Technik, Wirtschaft und Kultur\\Leipzig, Germany}
  \email{johannes.waldmann@htwk-leipzig.de}
}
\def\titlerunning{Efficient Completion of Weighted Automata}
\def\authorrunning{J. Waldmann}

\usepackage{amsmath}
\usepackage{amsfonts}
\newcommand{\ZZ}{\mathbb{Z}}
\newcommand{\BB}{\mathbb{B}}
\newcommand{\NN}{\mathbb{N}}
\newcommand{\FF}{\mathbb{F}}
\newcommand{\ola}[1]{\overleftarrow{#1}}
\newcommand{\ora}[1]{\overrightarrow{#1}}

\newtheorem{definition}{Definition}
\newtheorem{algorithm}{Algorithm}
\newtheorem{example}{Example}

\begin{document}
\maketitle
\begin{abstract}
  We consider directed graphs with edge labels from a semiring.
  We present an algorithm that allows efficient execution 
  of queries for existence and weights of paths, and allows updates of the graph:
  adding nodes and edges, and changing weights of existing edges.

  We apply this method in the construction of matchbound certificates
  for automatically proving termination of string rewriting.
  We re-implement the decomposition/completion algorithm
  of Endrullis et al. (2006) in our framework,
  and achieve comparable performance.
\end{abstract}

\section{Introduction}\label{sec:intro}

Our research is motivated by the efficient implementation of a completion algorithm
on finite weighted automata. Such an automaton is an edge-labelled
directed graph, and the algorithm requires to apply rules
of the form ``if some path does (not) exist, then add some nodes and edges'' repeatedly. 
In particular, we want to compute
a matchbound certificate for termination of string rewriting.

We are interested in an incremental algorithm.
First, it should allow to answer the question ``what is a cheapest path 
between these two nodes'' quickly.
From this information, the rules for completion
determine which nodes and edges to add, or weights to increase.
Then, after additions are performed, internal structures should be updated
so that further queries can be processed quickly.

We use the concept of weighted relation. A $S$-weighted relation $r$
from a set $P$ to a set $Q$ is a mapping $r : P\times Q\to S$.
In a dense representation, $r$ is a matrix that is indexed by $P\times Q$
with entries from $S$. In a sparse representation, 
$r$ is a directed bipartite graph (the partition has classes $P$ and $Q$) 
with edge labels from $S\setminus\{0\}$, where a missing edge represents weight $0\in S$.
If $S$ is a semiring, then $S$-weighted relations form a semiring as well.
We will recall notation for semirings and relations in Section~\ref{sec:waut}
and present basic implementation choices in Section~\ref{sec:rel}.

Section~\ref{sec:inc} presents the core of our approach:
efficient (re-)computation 
of products of relations after updates of edges.
Section~\ref{sec:cert} introduces the application area of matchbound
certificate construction. In Section~\ref{sec:semi},
we give an enhancement of the (max,min) semiring
that allows to apply the general algorithm for this application.

Algorithms are presented in a purely functional setting
and indeed we have implemented them in Haskell.
We obtain concise code where correctness is easy to see,
and should be straightforward to prove.
Performance that is comparable 
to an earlier purpose-built imperative implementation, 
see Section~\ref{sec:discussion}.

\emph{Acknowledgments.} A preliminary version of this paper
was presented at the TERMGRAPH 2016 workshop.
I thank reviewers and participants 
for critical reading and interesting discussion.
 
\section{Labelled Graphs for Weighted Relations and Automata}\label{sec:waut}

We present notation for standard concepts that will be used later on.

A \emph{semiring} $(S,+_S,\cdot_S,0_S,1_S)$ is a set $S$ with binary operations
$+_S$ (addition) and $\cdot_S$ (multiplication) such that $(S,+_S,0)$ is a commutative monoid,
$(S,\cdot,1_S)$ is a monoid, $0\cdot x=0=x\cdot 0$,
and addition distributes over multiplication from both sides.
We omit index $S$ if it can be inferred from context.
The natural numbers form a semiring $(\NN,+,\cdot,0,1)$ with addition and multiplication
in the standard sense.
A semiring is \emph{idempotent} if $x+x=x$. 
Examples for idempotent semirings are the \emph{Boolean} semiring $\BB=(\{0,1\},\vee,\wedge,0,1)$,
and the \emph{fuzzy} semiring $\FF=(\{-\infty\}\cup\ZZ\cup\{+\infty\},\max,\min,-\infty,+\infty)$.
In our setting, the Boolean semiring gives the usual (non-weighted) semantics
for relations and automata. The fuzzy semiring is used in the application area of matchbounds.

A $S$-\emph{weighted relation} $r$ from a set $P$ to a set $Q$ is a mapping $r:P\times Q\to S$.
Relations from $Q$ to $Q$ form a semi-ring with addition $(r_1+r_2)(p,q)=r_1(p,q)+_S r_2(p,q)$
and \[(r_1\cdot r_2)(p,q) = \sum_S \{ r_1(p,m)\cdot_S r_2(m,q) \mid m\in Q\}.\]
This corresponds to addition and multiplication of matrices over the semiring.
A Boolean-weighted relation is a relation in the usual sense,
where the weight $r(p,q)$ indicates whether $(p,q)$ is in the relation $r$ or not.

A $S$-\emph{weighted automaton}~\cite{Droste:2009:HWA:1667106} 
over an alphabet $\Sigma$ is a $\Sigma$-indexed collection
of $S$-weighted relations $A:\Sigma\to(Q\times Q\to S)$.
This function is extended from words to strings by multiplication of relations,
giving $A:\Sigma^*\to(Q\times Q\to S)$. 
The weight of a path is the product of the weight of its edges,
and $A(w)(p,q)$ is the sum of the weights of all $w$-labelled paths from $p$ to $q$.
Usually, the model also includes
projection functions (initial and final weights) but we do not need this.
Our automata form a monoid w.r.t. addition defined by $(A_1+A_2)(w)=A_1(w)+A_2(w)$.
A Boolean-weighted automaton is an automaton in the usual sense.

\section{Representing Weighted Relations}\label{sec:rel}

For the relations in our main application area (matchbounds), 
we expect that most nodes will have low degree, so we choose a
sparse representation.
We assume some efficient implementation of finite maps,
as given by \verb|Data.Map| (balanced search tree) 
or \verb|Data.IntMap| (Patricia tree) from the \verb|containers| package for Haskell
\url{http://hackage.haskell.org/package/containers}.
These provide fast individual queries, 
and also optimised bulk operations.

A weighted relation $r:P\times Q\to S$
could literally be represented as \verb|Map (p,q) s|,
with the provision that the map only holds pairs with nonzero weight.
This is inefficient since we often need to get all (nonzero) successors 
of a point $x\in P$ quickly.
This suggests \verb|r :: Map p (Map q s)|.
Then we can find successors of \verb|x| as \verb|keysSet (r M.! x)|.
But now the implementation is biased: 
we cannot easily find predecessors - which we do need, for efficient multiplication.
So, we also keep the representation in the opposite direction,
and a relation is a pair of nested maps:
\begin{verbatim}
data Rel p q s = Rel { fore :: Map p (Map q s)
                     , back :: Map q (Map p s) }
\end{verbatim}

A weighted automaton is simply represented as
\begin{verbatim}
data Aut q sigma s = Aut (Map sigma (Rel q q s))
\end{verbatim}
This model will be enhanced in Section~\ref{sec:inc}.

Our representation for weighted relations is similar to the 
representation of edge-labelled directed graphs in the \verb|fgl| 
library~\cite{DBLP:journals/jfp/Erwig01}.
\begin{verbatim}
type GraphRep a b = Map Int (Context' a b)
type Context' a b = (Map Int [b], a, Map Int [b])
\end{verbatim}
Here, \verb|Int| is the node type, \verb|a| is the node label type, \verb|b| is the edge label type.
The similarity is that two efficient maps are nested.
One difference is that \verb|fgl| graphs 
can have multiple edges between one pair of nodes (see \verb|[b]|),
while we just need one. This difference is not essential. 
We do not need multiple edges since we can add their weights in the semiring.
Another difference is that \verb|fgl| uses just one pair of maps, while we have two.
This allows \verb|fgl| to express the idea of \verb|Context|.
We think that our representation gives a better separation of concerns.
The difference in performance should be small, but we have not measured it.

We indicate how relations are multiplied. This feature is not present in \verb|fgl|,
presumably since it requires a semiring structure on the edge labels,
which is outside that library's scope.

We use a type class to express the signature for semirings:
\begin{verbatim}
class Semiring s where
  zero :: s ; one :: s 
  plus :: s ->  s -> s ; times :: s -> s -> s
\end{verbatim}
and derive an instance for relations:
\begin{verbatim}
instance Semiring s => Semiring (Rel q q s) where
  plus = plusWith plus ; times = timesWith plus times

plusWith 
    :: (s -> s -> s) -> Rel p q s -> Rel p q s -> Rel p q s
plusWith f r s = 
    Rel { fore = M.unionWith (M.unionWith f) (fore r)(fore s)
        , back = M.unionWith (M.unionWith f) (back r)(back s) }

timesWith 
    :: (u -> u -> u) -> (s -> t -> u) -> Rel p q s -> Rel q r t -> Rel p r u
timesWith f g = M.foldl (plusWith f) empty
              $ M.intersectionWith (combine f g) (back r) (fore s)

combine 
    :: (c -> c -> c) -> (a -> b -> c) -> M.Map p a -> M.Map q b -> Rel p q c
combine f g qp qr = 
    Rel { fore = M.map ( \ w1 -> M.map (\w2 -> f w1 w2) qr ) qp
        , back = M.map ( \ w2 -> M.map (\w1 -> f w1 w2) qp ) qr }
\end{verbatim}
The performance critical part is the combination of maps with these library functions:
\begin{verbatim}
M.unionWith        :: (a -> a -> a) -> Map q a -> Map q a -> Map q a 
M.intersectionWith :: (a -> b -> c) -> Map q a -> Map q b -> Map q c 
\end{verbatim}
Their performance to combine maps $m_1,m_2$ is $O(|m_1|+|m_2|)$.
We actually expect both functions to work in
in $O(\min (|m_1|,|m_2|) \cdot \log \max(|m_1|,|m_2|))$ time,
which is better when one of the maps is small.
Intuitively, this can be achieved by doing, for each element
of the smaller map, a lookup (or insertion, or deletion) in the larger map.
The actual implementation~\cite{Okasaki98fastmergeable} does something more clever
that detects the case when sub-trees have disjoint ranges.
We could not prove the conjectured performance bound from the library's source code,
but performance in examples confirms the conjecture.

\section{Queries and Updates}\label{sec:inc}

Given a weighted automaton 
(a $\Sigma$-indexed collection of $S$-weighted relations over $Q$),
we want to run queries $A(w)(p,q)$.

From the structure of our application, 
we know that $w$ is an element of a fixed set of query strings $W$,
and we will run the query for automata $A_i$, where $A=A_0\subseteq A_1\subseteq \ldots$
and in each step of that chain, just a few edges have been added.
We pre-process $W$ in order to store helpful extra data with the automata.

A first idea is to compute the relation $A(w)$
as the product of relations $A(w_1)\cdot\ldots\cdot A(w_n)$.
Given $A(w)$, each query $A(w)(p,q)$ reduces to a look-up, which is fast.

A second idea is to use relations between members of $W$,
to re-use sub-products computations. 
For example, if $\Sigma=\{b,c\}$ and $W=\{bbb,bbc\}$,
we should compute an intermediate relation $h=A(b)\cdot A(b)$, 
and then $A(bbb)=h\cdot A(b), A(bbc)=h\cdot A(c)$.
This is generalized as follows.

\begin{definition}
  A \emph{multiplication chain} for a finite set $W\subseteq\Sigma^*$
  is a finite set $C\subseteq \Sigma^+$ with 
  \begin{itemize}
  \item $\Sigma\subseteq W$ (the chain contains all one-letter words),
  \item for each $w\in W$ with $|w|>1$, there are $w_1,w_2\in C$ with $w=w_1\cdot w_2$,
  \item $W\subseteq C$.
  \end{itemize}
  The \emph{cost} of a chain $C$ is the number of words of length $>1$ in $C$
  (which is $|C|-|\Sigma|$).
\end{definition}
The cost of $C$ is the number of multiplications that are carried out
when all elements of $C$ are computed according to the decomposition.
For $W=\{bbb,bbc\}$, we have a multiplication chain $C=\{b,c,bb,bbb,bbc\}$
of cost 3. Finding an optimal multiplication chain for given $W$ 
looks like a hard combinatorial problem~\cite{DBLP:conf/soda/LehmanS02}.
As a fast approximation,
we use the idea of the RePair algorithm \cite{DBLP:conf/dcc/LarssonM99}
to repeatedly extract a most frequent pair of letters.
The pairings are modelled by
\begin{verbatim}
data Letter sigma = Unit sigma | Times (Letter sigma) (Letter sigma)
\end{verbatim}
and the enhanced automaton then has type
\begin{verbatim}
data Aut q sigma s = Aut (Map (Letter sigma) (Rel q q s))
\end{verbatim}
Having a good chain is essential for performance
since the cost for multiplication of relations is high
once the graph gets large.
It might pay to spend more preprocessing time
to obtain a better chain.

The third idea for efficient implementation is related to the
incremental nature of the application: nodes and edges are added.
So we have a chain of automata $A_0\subseteq A_1 \subseteq A_2 \subseteq\dots$
Consider one element $A=A_i$, and its successor $A_{i+1}=A'$.
We have $A' = A + \Delta$, and the typical case is that 
$A$ is large but $\Delta$ is small. 

Consider some $w$ from the multiplication chain,
with decomposition $w=w_1\cdot w_2$.
Then $A(w)=A(w_1)\cdot A(w_2)$,
and $A'(w) = (A(w_1)+\Delta(w_1))\cdot(A(w_2)+\Delta(w_2))$,
so $A'(w)=A(w) +  \Delta(w_1) \cdot A(w_2) + A(w_1)\cdot \Delta(w_2)+\Delta(w_1)\cdot\Delta(w_2)$.

\begin{algorithm}\label{alg:inc}
Doing this computation bottom-up along the multiplication chain (towards increasing lengths),
we can compute $A'(w)$ for each $w$ in the chain.
\end{algorithm}

Note that in each multiplication,
at least one of the arguments is $\Delta(\cdot)$, that is, small.
Then also the product is small, implying that in each sum,
one summand is small. This matches the discussion at end of Section~\ref{sec:rel}.

\section{Matchbound Certificates}\label{sec:cert}

We now come to the application-specific part of the paper.

The matchbound method~\cite{DBLP:journals/aaecc/GeserHW04}
proves termination of string rewriting
(and recently, of cycle rewriting~\cite{DBLP:conf/rta/ZantemaKB14}).
The method can be presented via automata
with weights in the fuzzy semiring $\FF$~\cite{DBLP:journals/jalc/Waldmann07}.

An $\FF$-weighted automaton $A$ is \emph{compatible} 
with a string rewriting system $R$ over $\Sigma$
if for each $(l,r)\in R$, and $p,q\in Q$, we have that $A(l)(p,q)<_0A(r)(p,q)$
where $<_0$ denotes the relation on $\FF$ defined by $x<_0y$ iff $x<y \vee x=0_\FF=y$.

An automaton  that contains a $\Sigma$-\emph{flower} 
(a state $p$ with transitions $p\stackrel{c:0}{\to} p$ for each $c\in\Sigma$)
and that is compatible with $R$, is called a \emph{matchbound certificate}
for termination of $R$. This concept is extended to RFC-matchbounds,
but this is orthogonal to the present paper, so we do not discuss this extension.

For constructing matchbound certificates, 
one starts with the flower graph and adds paths whenever
compatibility is violated.
There is a method~\cite{DBLP:journals/ijfcs/GeserHWZ05}
that uses heuristics to re-use nodes when adding paths,
but this may fail to terminate even if a certificate exists.
There is a complete method~\cite{decomp} based on the idea
of extending the alphabet by formal left and right inverses of letters.
We focus on that method.
 
We introduce epsilon transitions. 
In the graph model, these are unlabelled edges.
In the automaton model, we add a $Q$-by-$Q$ matrix $A_\epsilon$
where each entry is $0_\FF$ (no edge) or $1_\FF$ (epsilon edge).
We denote $A_\epsilon(w) := A_\epsilon A(w_1) A_\epsilon \cdot \dots \cdot 
A_\epsilon A(w_n) A_\epsilon$.
Rules given below will ensure that (ultimately)
$A_\epsilon$ is reflexively and transitively closed.

We extend the alphabet. For each $c\in \Sigma$,
we introduce fresh letters $\ola{c}$, and $\ora{c}$.
These will act as formal right and left inverses.
We define $\ola{w}=\ola{w_n}\cdot\dots\cdot\ola{w_1}$ 
and $\ora{w}=\ora{w_n}\cdot\dots\cdot\ora{w_1}$ .

\begin{algorithm}\label{alg:mb}
To produce a matchbound certificate for a string rewriting system $R$ 
over alphabet $\Sigma$,
start with the $\Sigma$-flower automaton $A$,
and $A_\epsilon$ as the identity relation,
and repeatedly apply the following rules:
\begin{enumerate}
\item {} [TRANSITIVE]
  if there are $p,q$ with $(A_\epsilon \cdot A_\epsilon)(p,q)\neq 0_\FF$, 
  add epsilon transition from $p$ to $q$.
\item {} [INVERSE]
  if there are $c,p,p',q',q$ with $A_\epsilon(p',q')=1_\FF$ and \\
  $A(c)(p,p')\ge A(\ola{c})(q',q)>0_\FF $ or $0_\FF<A(\ora{c})(p,p')\le A(c)(q',q),$\\
  add epsilon transition from $p$ to $q$. 
\item {} [REWRITE]
  if there is $(l,r)\in R$ such that there is $(p,q)$
  such that $A_\epsilon(l)(p,q)\not <_0 A_\epsilon(r)(p,q)$,  then:
  \begin{itemize}
  \item let $p'\stackrel{c:h}{\to} q'$ with $c\in\Sigma,h\in\FF$ 
    be a transition of minimal height on a maximal $l$-labelled $(p,q)$-Path,
    such that $l=s c t$ for $s,t\in\Sigma^*$,
  \item then add a path from $p'$ to $q'$ over fresh states only
    that consists of a sequence of edges labelled by
    $\ora{s}$ with height $h$, $r$ with weight $h+1$, $\ola{t}$ with height $h$.
  \end{itemize}
\end{enumerate}
Rule [REWRITE] should only be applied 
if none of [TRANSITIVE] and [INVERSE] is applicable.
The algorithm stops when no rule applies.
\end{algorithm}

\begin{example}
For the $R=\{aa\to aba\}$ over $\Sigma=\{a,b\}$,
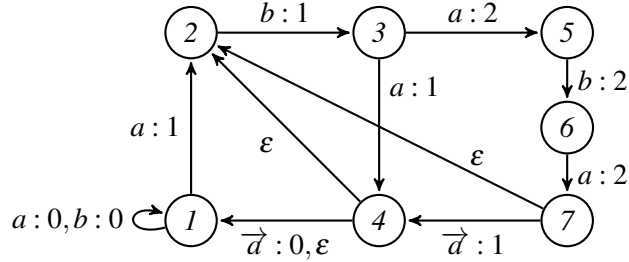
\begin{center}
\begin{tikzpicture}[->,>=stealth',shorten >=1pt,auto,node distance=2.5cm,
  thick,
  main node/.style={
  draw,circle%
  }
]

\node[main node](1){1} ;
\node[main node](2)[above of=1]{2} ;
\node[main node](3)[right of=2]{3} ;
\node[main node](4)[right of=1]{4} ;
\node[main node](5)[right of=3]{5} ;
\node[main node](6)[below=0.55cm of 5]{6} ;
\node[main node](7)[right of=4]{7} ;

\path 
(1) edge [loop left] node {$a:0, b:0$} (1)
(1) edge node {$a:1$} (2) 
(2) edge node {$b:1$} (3)
(3) edge node [pos=0.2] {$a:1$} (4) 
(4) edge node {$\ora{a}:0, \epsilon$} (1)
(3) edge node {$a:2$} (5)
(5) edge node {$b:2$} (6)
(6) edge node {$a:2$} (7)
(7) edge node {$\ora{a}:1$} (4)
(4) edge node {$\epsilon$} (2)
(7) edge [above] node [ pos=0.2 ] {$\epsilon$} (2)
;

\end{tikzpicture}

\end{center}
Weights of $\epsilon$ edges are $1_\FF$.
Reflexive $\epsilon$ edges (loops) are not shown.
We start with the $\{a_0,b_0\}$-flower at state $1$.
We apply rule [REWRITE] (for $p=p'=q'=q=1$), producing the path 
$1\stackrel{a:1}{\to} 
2\stackrel{b:1}{\to} 
3\stackrel{a:1}{\to} 
4\stackrel{\ora{a}:0}{\to} 
1$, then [INVERSE] (for $p=4,p'=1=q'=q$, and for $p=4,p'=1=q', q=2$), 
producing $4\stackrel{\epsilon}{\to} 1$
and $4\stackrel{\epsilon}{\to} 2$,
then [REWRITE] (for $p=p'=3, q'=4,q=2$), producing the path
$3\stackrel{a:2}{\to} 
5\stackrel{b:2}{\to} 
6\stackrel{a:2}{\to} 
7\stackrel{\ora{a}:1}{\to} 
4$,
then [INVERSE] (for $p=7,p'=4,q'=1,q=2$), 
producing $7\stackrel{\epsilon}{\to} 2$,
\end{example}

It has been shown that this construction terminates 
iff $R$ is matchbounded. 
In the example, the matchbound (the highest label) is 2.
An efficient implementation had been given by Endrullis.
It can build matchbound certificates with tens of thousands of states 
in a few seconds. 
The implementation was purpose-built, using an imperative programming style, in Java.

\section{An Extension of the (max,min) Semiring}\label{sec:semi}

To realize matchbound certificate construction  (Algorithm~\ref{alg:mb})
in the general framework (Algorithm~\ref{alg:inc}),
we enrich edge labels (weights) with extra information.

As a motivation, consider rule [REWRITE]. With weights from $\FF$,
we can immediately check the condition $A_\epsilon(l)(p,q)<_0 A_\epsilon(r)(p,q)$
by two lookups. For the case that we have to add a path, 
we need a minimal edge $p'\stackrel{c:h}{\to}q'$.
We do not want to compute that information on the spot,
but have it available already in the weight $A_\epsilon(l)(p,q)$.

In the present section, we will handle
\begin{itemize}
\item for rule [TRANS]: $\epsilon$ transitions (their weights will be $1_\FF$),
\item for rule [INVERSE]: formal left and right inverses (we introduce inverse elements for $\FF$),
\item for rule [REWRITE]: the location of a minimal edge (we introduce position information).
\end{itemize}
The challenge is to do this in a way that still allows a semiring structure.

To model epsilon transitions,
we extend the alphabet by a fresh letter: $\Sigma_\lambda=\Sigma\cup\{\lambda\}$
with the intention that $A(\lambda)$ is the $\epsilon$ transition relation.
Is weights are $0_\FF$ (no edge) and $1_\FF$ (edge).
Then, for a word $w\in\Sigma^*$ with letters $w_1w_2\dots w_n$ 
we define $w_\lambda = \lambda w_1 \lambda w_2 \lambda \dots \lambda w_n \lambda\in\Sigma_\lambda^*$.
The query set (see Section~\ref{sec:inc}) contains
\begin{itemize}
\item $\{\lambda \lambda\}$ because of rule [TRANSITIVE],
\item $\{\ora{c} \lambda c , c\lambda \ola{c} \mid c\in\Sigma \}$ for [INVERSE],
\item $\{ l_\lambda , r_\lambda \mid (l,r)\in R \}$ for [REWRITE].
\end{itemize}

To realize rule [TRANSITIVE],
we compare $A(\lambda \lambda)$ to $A(\lambda)$
and add all edges that are present on the left, but not on the right.

To realize rule [INVERSE],
we want to compare $A(\ora{c} \lambda c)$ (and $A(c\lambda\ola{c})$, respectively)
to $A(\lambda)$.
We need to realize the check for $A(\ora{c})(p,p')\le A(c)(q',q)$.
To this end, we extend the semiring domain by left and right inverted numbers 
$\ora{f}, \ola{f}$ for $f\in\NN$,
with multiplication rules 
\[ \ora{f}\cdot g = \text{if}~ f\le g ~ \text{then} ~ 1_\FF ~ \text{else} ~ 0_\FF.
\]
Then the product  $A(\ora{c} \lambda c)$  will contain only $0_\FF$ and $1_\FF$.

The algorithm requires to add a path $\ola{s}r\ora{t}$
under certain conditions. To find the start and end points of
this path, we need to find an edge of minimal weight,
among all maximal $l$-paths from $p$ to $q$.
We already have the weight of that edge --- it is just the value
in the matrix $A(l)$ (interpretation of $l$) at position $(p,q)$.
But we lost the information on where that minimal edge is located:
we need both the location in the automaton (start and end node of the edge)
and the location in the string (number of letters before that edge).
So we carry along the following extra information
\begin{verbatim}
data I = I { weight :: F -- ^ the original information
           , from  :: Q, to :: Q -- ^ start and end of edge with that weight
           , offset :: Int, total :: Int
           }
\end{verbatim}
and these semiring operations
\begin{verbatim}
plus  i j = if weight i <= weight j then i else j -- ^ min operation
times i j = if weight i >= weight j -- ^ max operation, update offsets
  then i { total = total i + total j }
  else j { total = total i + total j , offset = total i + offset j }
\end{verbatim}
where we use the Haskell notation of ``record update'' \verb|base { name = val }|
where components that are not mentioned in braces, have their value from \verb|base|.
So, when we have $A(l)(p,q)=i$, and need to find the decomposition into $l=sct$,
where $c$ has minimal weight,
we determine $s$ as the prefix of length \verb|offset i|,
and the $c$ edge is \verb|(from i, to i)|.

This construction satisfies semiring axioms only up to some equivalence relation
because the choice of result in case \verb|weight i == weight j| is arbitrary.

\section{Summary and Discussion}\label{sec:discussion}

The contributions of this paper are
\begin{itemize}
\item 
  a general method for efficient path queries under edge updates,
  see Section~\ref{sec:inc}.
\item 
  a specific enhancement of the fuzzy semiring, 
  to implement matchbound certificate construction in this general setting,
  see Section~\ref{sec:semi},
\end{itemize}
These methods have been used 
in a recent re-implementation of Endrullis' algorithm
in the 2016 version 
(\url{https://gitlab.imn.htwk-leipzig.de/waldmann/pure-matchbox})
of the Matchbox termination prover.
It exclusively uses RFC matchbound for standard termination, 
and matchbounds for cycle termination.
Performance is in the same ballpark as the original implementation.
An RFC matchbound certificate for \verb|SRS/secret06/jambox1| 
(with 43495 states, for height 12
 --- this is the ``killer example'' in \cite{decomp}) 
is constructed in 8 seconds on a standard desktop computer.

While we update monotonically (add edges, or increase weight of edges),
Algorithm~\ref{alg:inc} could as well be used for deletions---if the weight
domain is a ring, that is, provides subtraction as well.

For automatically proving termination,
it would be interesting to apply our method for automata completion
in other semirings,
e.g., for proving termination 
with matrix interpretations over the natural or arctic numbers.
Known implementations use small~\cite{DBLP:conf/rta/HofbauerW06}
or medium sized~\cite{munomu} graphs.
Using our method, larger graphs could be handled.
Still, we do not know any completeness result 
for constructing matrix interpretations over $\NN$ or $\mathbb{A}$ by completion,
and so far, there is no suitable notion of decomposition using formal inverses.

For graph rewriting,
we recommend matchbound certificate construction
as a test case for implementations. 

\bibliographystyle{plainurl}

\end{document}